\documentclass[11pt]{article}
\usepackage[utf8]{inputenc}
\usepackage[T1]{fontenc}
\usepackage{lmodern}
\usepackage{authblk}
\usepackage{mdframed}
\usepackage{amsmath, amssymb, amsthm}
\usepackage{graphicx}
\usepackage{booktabs}
\usepackage{float}
\usepackage[margin=1in]{geometry}
\usepackage{placeins}
\usepackage{enumitem}
\usepackage{fancyhdr}
\usepackage{caption}
\usepackage{hyperref}
\hypersetup{hidelinks}
\usepackage[backend=biber,style=numeric,sorting=none,defernumbers=true]{biblatex}
\title{\textbf{Promoting AI Literacy in Higher Education: Evaluating the IEC-V1 Chatbot for Personalized Learning and Educational Equity}}
\author[1]{Dr. Stefan Pietrusky}
\affil[1]{Down Church Studios, Hintergasse 46a, 67150 Niederkirchen, Germany}
\affil[1]{Heidelberg University of Education}
\affil[1]{\texttt{pietrusky@downchurch.studio}}
\date{\today}

\makeatletter
\let\blx@rerun@biber\relax
\makeatother

\begin{document}

\maketitle

\begin{mdframed}[linewidth=0.5pt, linecolor=black, frametitle={Abstract}, frametitlealignment=\centering]
\noindent
The unequal distribution of educational opportunities carries the risk of having a long-term negative impact on general social peace, a country's economy and basic democratic structures. In contrast to this observable development is the rapid technological progress in the field of artificial intelligence (AI). Progress makes it possible to solve various problems in the field of education. In order to effectively exploit the advantages that arise from the use of AI, prospective teacher training students need appropriate AI skills, which must already be taught during their studies. In a first step, the added value of this technology will be demonstrated using a concrete example. This article is therefore about conducting an exploratory pilot study to test the Individual Educational Chatbot (IEC-V1) prototype, in which the levels can be individually determined in order to generate appropriate answers depending on the requirements. The results show that this is an important function for prospective teachers and that there is great interest in taking a closer look at this technology in order to be able to better support learners in the future. The data show that experience has already been gained with chatbots, but that there is still room for improvement. It also shows that IEC-V1 is already working well. The knowledge gained will be used for the further development of the prototype to further improve the usability of the chatbot. In general, it is shown that useful AI applications can be effectively integrated into learning situations even without proprietary systems and that important data protection requirements can be met.
\end{mdframed}

\section{Introduction}
In the context of education and learning, the use of AI will make a significant contribution to achieving often discussed societal and political goals, such as the Sustainable Development Goals (SDGs) \cite{1} \cite{2} \cite{3}
\cite{4}. This is made possible by the creation of a system that provides everyone with the same or the best opportunities to learn successfully, depending on their individual prerequisites. This was described long before the current developments by the Pearson Learning Research Institute, among others \cite{5}. At the same time, it is important to prevent the technological advances of recent years, which are now publicly perceived, from leading to systems that are based only on algorithms and AI and, without control, do not promote more fairness, but possibly social inequalities in the education system \cite{6} \cite{7}, i.e. strengthen the digital divide \cite{8}. One example is the White Paper adopted by the European Commission, in which ethnic and social challenges that may arise from AI can be prevented through regulatory approaches \cite{9}. 

In the meantime, numerous teaching and learning tools have already been created, which are used by both teachers and students at some universities or in certain departments or institutes. How and whether tools are used depends on one's attitude towards this technology, or user-friendliness and usefulness determine acceptance and use, according to the TAM model \cite{10}. The use also depends on the costs incurred. Many AI applications are proprietary and use freemium models \cite{11}, which means that the actual function is only available to a limited extent for testing. For full use, a bank account must then be stored. Freemium models are also necessary in this area, as large computing power must be available to process requests quickly and at a certain quality. In this context, there is already a discussion about the effects of energy consumption and CO2 emissions on the environment through the various AI applications and their models \cite{12}. 

The foundations of numerous start-ups in the AI field have become possible due to university research. Nevertheless, it often happens that the state, i.e. public tax funds and other public funding, support the development of technological innovations that are then used in private-sector applications \cite{13} \cite{14}. Of course, there are also models that have been developed directly by university and state funding or directly by companies and are available via open source (e.g. object recognition YOLO11 from Ultralytics, language conversion EdgeTTS from Microsoft, text comprehension Llama3.2 from Meta, etc.). These models can be used to conduct your own AI experiments without having to rely on large data centers. In various model management and delivery systems such as Hugging Face, Ollama, TensorFlow Serving, ONNX Runtime, MLflow, etc., models are also available for free use at times, which perform better than paid applications in certain areas. The problem is that the use of these freely usable models requires a technical understanding, certain computing power and knowledge of their existence in general in order to be able to use them in learning situations. These requirements can be summarized under the term AI literacy.

If you don't have the specific expertise, you'll automatically turn to commercial AI apps. Due to this fact and the financial requirements described, there is again a risk that those who can afford to use proprietary applications to bypass the technical setup process and then there are those who cannot afford to use AI apps and are excluded from the benefits of this technology due to the mentioned requirements. To prevent this, it is important to teach basic AI skills not only to prospective teachers, but also to teachers already working in schools, so as not to further exacerbate existing inequalities. The aim is to show teachers how to set up their own chatbots locally, how to use them in the classroom and how to make them available to learners at certain time slots. 

In a first step, this is to be made possible with the help of the prototype IEC-V1. IEC-V1 differs from other applications (e.g. ChatGPT from OpenAI) because learners can individually determine the level at which answers should be generated. Since the chatbot's knowledge base is also individually configurable, the IEC-V1 can solve another problem that has not yet been researched much. There is already research whose data shows that teachers and learners rely too much on the output of AI apps without questioning their accuracy, which influences the development of important skills such as critical thinking or self-regulation \cite{15}. When AI is used for learning, the user does not know whether what the machine writes is correct, especially if there is no one available to check whether the output is correct. The fact that large language models (e.g. GPT) reproduce information without specifically checking whether it is correct has already been criticized by studies \cite{16}. To prevent this, IEC-V1 allows the sources, i.e. the knowledge database to which the LLM accesses, to be defined by the users themselves. This illustrates the relevance of the actual control of sources, which is used as a basis by an LLM. In a specific learning situation, the teacher would then specify which sources the learners are allowed to use to work on tasks. 

As part of the first testing of the prototype, prospective teachers of a seminar at the Heidelberg University of Education were able to test the IEC-V1 at the Institute for Foreign Languages and evaluate its user-friendliness. Previous experiences with chatbots were also recorded. The results of the survey are discussed in the chapter of the same name.  The following chapter is about setting up the application. 

\section{Materials and methods}
Setting up a chatbot is now possible in different ways. There are now various platforms that can be used to create bots without having to program yourself. An example of this is Botpress, which, like WordPress, is designed as a modular system to simplify the setup \cite{17}. It's available as an open-source project on GitHub, but it's also offered with advanced features as a commercial solution. Various tools were used for the development of IEC-V1 in order to be as flexible as possible and not to have any limitation in terms of the possible uses that result from freemium models.  

First, a Large Language Model (LLM) was selected, i.e. a model that was trained to understand, generate and respond to natural language. On the open source platform Ollama, several language models (Mistral, Gemma, Llama, etc.) are available that can be executed locally. The advantage here is that no data is sent to external servers due to the local execution, which is often a problem for many apps when it comes to use in schools. Ollama thus offers decisive added value compared to proprietary applications, which often store data on servers abroad and whose processing is therefore not covered by the European Union's GDPR. With some Ollama models, you can also adjust the parameters (e.g. temperature) individually via a model file in order to better regulate the output by reducing creativity. There is no cost to use Ollama and the system is easily installable on various operating systems via the terminal. After Ollama was installed, a suitable LLM, specifically Llama3.1 from Meta with 8B parameters, was selected. There are other variants of this LLM (70b and 405b) with different quantization levels available, but since the IEC-V1 should be usable in as many situations as possible, a variant was selected that works without major requirements in terms of required computing power. The use of Llama3.1 in the context of IEC-V1 is essentially limited to the text comprehension capability of the model. The knowledge base is defined by the respective users by entering URLs and PDF files, which means that the knowledge on which the model was trained does not matter.  

\begin{figure}[htbp]
    \centering
    \includegraphics[width=1.0\textwidth]{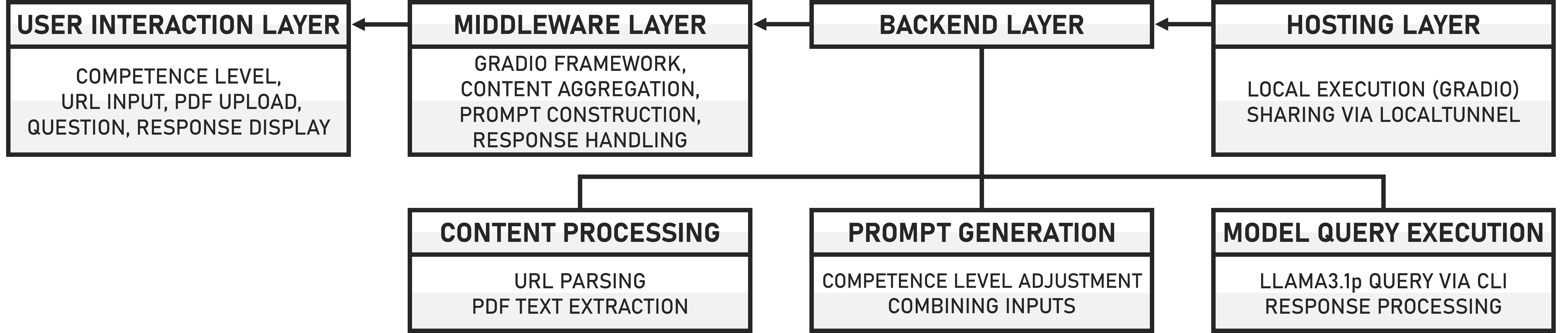}
    \caption{Dependency-Focused System Architecture of IEC-VI Chatbot.}
    \label{F1}
\end{figure}

In order to interact with the chatbot, a user interface had to be created. Various options were available here (e.g. Streamlit, Gradio, Anvil, Django, etc.). Initially, it was planned to implement the interface itself via HTML, CSS and JavaScript. However, since the goal is to integrate the application as easily as possible in the context of school, the framework of Gradio was used to create the interactive web application. This reduced the complexity of the code. Due to the declarative programming method, an initial implementation with the help of Streamlit led to user inputs, such as the adjustment of the levels, automatically generating new outputs of the LLM, despite integrated caching mechanisms. Due to the statelessness of the streamlit scripts, the interface of IEC-V1 was implemented with the help of Gradio. A dependency-oriented representation of the system architecture for the IEC-V1 prototype is presented below (see Fig.~\ref{F1}). 

\pagebreak

The illustration highlights the hierarchical dependencies between the user interaction, middleware, backend and hosting levels. The data flow and responsibilities are represented by the arrows. The user first selects the skill level (Beginner, Intermediate or Advanced) and then enters the URL. If there are several, they are separated by a comma. Alternatively, PDF files can be uploaded or both source types can be defined. When the source is defined, a question is entered and submitted to the LLM, after which a response is issued. 
For the actual programming of the application, the code editor Visual Studio Code (VS-Code) from Microsoft was used. VS-Code was chosen because it supports many programming languages and it was initially planned to implement the IEC-V1 user interface itself. 

After testing how IEC-V1 works, the next step was to find a way to make the chatbot available as part of an initial trial without users having to install all dependencies on their own devices. 
Both Gradio and Streamlit offer the possibility to make applications available to others via a link. With Streamlit, a network URL is generated when you start the application. If the server and client are on the same network, they can be used to interact with the app. Since the implementation was done with Gradio, the automatically generated public URL could have been used. As part of the test, due to sudden problems with the firewall of the server computer, a tunnel was created using LocalTunnel. This made it possible to communicate between the local server and the students' end devices via the Internet. The individual IP addresses of the students' devices served as the password for the application.

The IEC-V1 was tested as part of a seminar on the use of digital media in foreign language teaching. After a short introduction by the actual seminar instructor and instructions on how to start the chatbot, the students were able to implement the theoretical content (URL and PDF files) in the application in order to ask the chatbot questions based on it. 
In order to verify the functioning of IEC-V1, an exploratory pilot study was conducted. The focus was on students' existing experience with chatbots in general and on the concrete evaluation of the user-friendliness of IEC-V1. The data was collected with the help of an online questionnaire, which could be accessed via a link in the chatbot's interface as part of the trial. The questionnaire was completed by the students on site in the seminar. Permission to conduct the study was obtained from the person responsible for the seminar and the institute management. The students were informed about their rights regarding the processing of the data. The declaration of consent was obtained by participating. The questionnaire contained closed questions that could be answered with a five-point Likert scale and an optional open-ended question. Here, the students were able to make suggestions for improvements. A total of 10 questions could be answered. The results of the study are discussed in the following chapter.

\section{Results}
Of the 9 students present, 7 completed the questionnaire in full. In the closed questions, a high value (5) means agreement and a low value (1) means rejection of the respective statement. The mean value for the students' previous experiences with chatbots was 3.29 with a dispersion or standard deviation of 1.38, which indicates a large dispersion in terms of the experience already gained. In terms of satisfaction with previous experiences with chatbots, the mean value is 3.14 (1.21), which indicates good satisfaction. In terms of ease of interaction with chatbots compared to other digital interaction options, the average score was 3.8. Again, there is a dispersion between students, as shown by the standard deviation of 0.98. 

In terms of user-friendliness, specifically related to IEC-V1, the interaction was rated as easy with a mean value of 3.86 (1.06). The functionality to customize the proficiency levels or the level of responses of IEC-V1 received a mean score of 4.14 with a low dispersion (0.69), highlighting the importance of this feature for students. The quality of the answers of the first prototype was rated with an average score of 3.29 (0.49), with a slight deviation, indicating a neural to slightly positive rating. The chatbot's response speed tended to be positive on average at 3.86 (1.06). The design of the IEC-V1 was rated with a mean score of 3.57 (1.27), indicating a positive perception. The open question, which involved general comments about the chatbot and suggestions for improvement, was answered by two students. On the one hand, the feedback was about the fact that a student's uploaded PDF file was not recognized and the waiting time for an answer from the chatbot took too long for some questions. The other feedback suggested that the outputs could be even more differentiated at the different levels in terms of general word choice, so a simpler vocabulary should be used. The results (see Table~\ref{t1}) of the exploratory pilot study show that the students have already had experience with chatbots.

\captionsetup{justification=centering}

\begin{table}[htbp]
\centering
\begin{tabular}{p{7cm} c c}
\toprule
\textbf{Ask} & \textbf{Mean} & \textbf{Standard deviation} \\ 
\midrule
Previous experience [experience] & 3.29 & 1.38 \\
Previous experience [satisfaction] & 3.14 & 1.21 \\
Previous experience [interaction] & 3.43 & 0.98 \\
Ease of Use [Interaction] & 3.86 & 1.07 \\
Ease of Use [Adjust Skill Levels] & 4.14 & 0.69 \\
Ease of Use [Response Quality] & 3.29 & 0.48 \\
Ease of Use [Speed of Response] & 3.86 & 1.07 \\
Usability [Design] & 3.57 & 1.27 \\
\bottomrule
\end{tabular}
\centering
\caption{Overview of questions about the user-friendliness of the IEC-V1 chatbot.\\
Mean values and standard deviations of the participant evaluations.}
\label{t1}
\end{table}

The evaluation of the data shows that the prototype was rated positively by the students in general. The function of individually determining the competence levels in order to influence the answers of the LLM was rated particularly positively. The data show that there is potential for optimization in terms of user-friendliness and technical reliability of the application. For example, the processing of PDF files must be improved, and the overall processing speed must be accelerated so that questions are answered faster.
The quality and differentiation of the answers, especially the choice of words of the chatbot, must be further adapted. After conducting the study, there was clear interest in further use of the chatbot. For example, the students specifically asked when IEC-V1 will be available and whether it can also be used independently of setting up their own local server. 

\section{Discussion}
As part of the pilot study, there was initially a problem with the firewall of the computer that was used as a server. As a result, the public URL automatically generated by Gradio could not be output. Specifically, access to a network socket was blocked due to a permission restriction. As a result, the less stable and performance-limited open source tool LocalTunnel had to be used. Although the students were able to access the server, the connection was not stable throughout and sometimes led to interruptions when processing input. Of the available 16 GB of VRAM, only a maximum of 6 GB was required, even with several requests on average. 

Two students in the seminar were unable to access the chatbot at all via their devices. This can probably be explained by a limitation of the connection established via LocalTunnel. To solve this problem in further testing, LocalTunnel is no longer used and instead either the firewall is temporarily disabled or exception rules are defined to enable certain ports so that the public URL automatically generated by Gradio can be used. Another option is to host the Gradio app via HuggingFace, which would allow the use of IEC-V1 without a local server. Other options, which would be costly to set up, are the use of Docker, Amazon Web Service (EC2 instance), Ngrok, Python Anywhere or, if Streamlit is used for the user interface, its cloud function.

The data collected from the pilot study and the resulting findings form the basis for the further development of the IEC-V1 prototype. Improving connection stability is a special priority here, as it has a major impact on the usability of the application. Regardless of this, the interaction with the chatbot will be expanded with further functions, which will have an impact on the quality of the answers and the speed of response at the same time. Specifically, the extraction of the text from the URL or PDF files is adjusted. It removes excess whitespace, line breaks, and distracting HTML elements. The extracted text is then bundled as continuous text and split into smaller blocks, which are then processed by the LLM. This means that all text is taken into account when processing questions and does not exceed the token limit of the Llama3.1 (7B) model. The option is also integrated that the text filtered from the sources can be displayed in the application if necessary, but can also be deleted. The extracted text, which serves as a knowledge base for the LLM to answer questions asked, is thus directly available and does not have to be reprocessed for each new question. This will have a positive effect on the overall reaction speed of IEC-V1, as well as on interaction as the process becomes clearer. The problem of PDF files not being recognized and processed should also be solved by this. Initial tests have shown that no more errors of this kind occurred after the adjustment. 

The described adjustments regarding the processing of the sources provided to the chatbot have a direct impact on the quality of the generated responses. As the overall context is now captured more precisely, the answers also become more concrete. In addition, the system messages of the selectable levels were to be adapted or expanded during further tests. This could also be defined directly by the participants or individually adapted as part of planned further studies. The adjustments would then have to be made either directly in the Python script of IEC-V1 or via an upstream interface in which various parameters can be defined. Replacing the LLM or switching from Llama3.1 to Llama3.2 or implementing your own models is also a way to see if the application delivers better results. 

Since the results of the pilot study show that the design of the IEC-V1 prototype did not appeal to all participants, the next version of the chatbot will integrate a way for the user to determine the visual representation himself. Gradio has an integrated design engine that can be used to change the appearance.
At the beginning, it was said that chatbot users rely too much on the generated results. To prevent the exchange with the chatbot without entering URL or PDF files, a corresponding condition has been integrated. In the adapted version, there must now be at least one source for a question to be asked. Checking the sources that are made available to learners is then the teacher's job. This procedure prevents the LLM from generating responses that may not be correct based on the dataset with which it was trained. Unfortunately, certain parameters can often lead to models fantasizing creatively. 

The exploratory pilot study carried out to test the IEC-1 prototype has provided the first important insights into the potentials that arise from the use of AI in the context of heterogeneous learning groups. With the appropriate functionality, chatbots offer the opportunity to support learning groups more individually and thus improve educational opportunities for everyone. 

\section{Conclusion}
In this article, it was clarified how the prototype of the IEC-V1 can help to accommodate different learning requirements by offering individually selectable levels, which are made possible by different system messages. It was shown that you are not dependent on proprietary applications for development, but can implement many things with the help of open source models and execute them locally. The results of the pilot study show that the prototype of the IEC-V1 has already been accepted by student teachers, and that there is an interest in using it in concrete learning situations or understanding the concrete functionality. 

In a next step, the knowledge gained will be used for the further development of the chatbot in order to improve the overall user-friendliness. In the future, prospective teachers from all disciplines should be more involved in the development of AI applications for education. The goal must be not only to provide the next generation of teachers with tools for passive use, but also to provide them with the skills to adapt and develop them independently depending on their learning situation in order to promote sustainable equal opportunities in the education system. 

\printbibliography[heading=bibintoc, notcategory=datasources, title={References}]

\end{document}